\documentclass[aps,twocolumn,superscriptaddress,nofootinbib,showpacs]{revtex4}

\usepackage{graphicx}
\usepackage{amssymb,amsmath,amsbsy}

\usepackage{psfrag}

\usepackage{epsfig}

\def\be{\begin{equation}}
\def\ee{\end{equation}}
\def\ba{\begin{eqnarray}}
\def\ea{\end{eqnarray}}

\begin{document}
%\begin{flushright}
%CFTP/09-033\\[-1mm]
%arXiv:????.???? [hep-ph]\\[-1mm]
%\end{flushright}
%\vspace*{1cm}

\title{Evading death by vacuum}
\author{A. Barroso}
\affiliation{Centro de F\'{\i}sica Te\'{o}rica e Computacional,
    Faculdade de Ci\^{e}ncias,
    Universidade de Lisboa,
    Av.\ Prof.\ Gama Pinto 2,
    1649-003 Lisboa, Portugal}
\author{P.M.~Ferreira}
    \email[E-mail: ]{ferreira@cii.fc.ul.pt}
\affiliation{Instituto Superior de Engenharia de Lisboa,
	1959-007 Lisboa, Portugal}
\affiliation{Centro de F\'{\i}sica Te\'{o}rica e Computacional,
    Faculdade de Ci\^{e}ncias,
    Universidade de Lisboa,
    Av.\ Prof.\ Gama Pinto 2,
    1649-003 Lisboa, Portugal}
\author{I.P.~Ivanov}
    \email[E-mail: ]{igor.ivanov@ulg.ac.be}
\affiliation{IFPA, Universit\'{e} de Li\'{e}ge, All\'{e}e du
6 Ao\^{u}t 17, b\^{a}timent B5a, 4000 Li\'{e}ge, Belgium}
\affiliation{Sobolev Institute of Mathematics, Koptyug avenue 4, 630090, Novosibirsk, Russia}
\author{Rui Santos}
    \email[E-mail: ]{rsantos@cii.fc.ul.pt}
\affiliation{Instituto Superior de Engenharia de Lisboa,
	1959-007 Lisboa, Portugal}
\affiliation{Centro de F\'{\i}sica Te\'{o}rica e Computacional,
    Faculdade de Ci\^{e}ncias,
    Universidade de Lisboa,
    Av.\ Prof.\ Gama Pinto 2,
    1649-003 Lisboa, Portugal}
\author{Jo\~{a}o P.~Silva}
    \email[E-mail: ]{jpsilva@cftp.ist.utl.pt}
\affiliation{Instituto Superior de Engenharia de Lisboa,
	1959-007 Lisboa, Portugal}
\affiliation{Centro de F\'{\i}sica Te\'{o}rica de Part\'{\i}culas (CFTP),
    Instituto Superior T\'{e}cnico, Universidade T\'{e}cnica de Lisboa,
    1049-001 Lisboa, Portugal}

\date{\today}

\begin{abstract}
In the Standard Model, the Higgs potential allows only one minimum at tree-level. 
But the open possibility that there might be two scalar doublets
enriches the vacuum structure, allowing for the risk that we might now be in a metastable state,
which we dub the panic vacuum. Current experiments at the LHC are probing the
Higgs particle predicted as a result of the spontaneous symmetry breaking.
Remarkably, in the two Higgs model with a softly broken $U(1)$ symmetry, the LHC experiments
already preclude panic vacuum solutions. 
\end{abstract}


\maketitle

The Standard Model (SM) of particle physics is a remarkably successful theory. 
Over the past forty years, the SM has passed numerous tests,
predicting, and agreeing with, experimental observables with great accuracy.
A crucial element of the Model is the mechanism of spontaneous symmetry breaking (SSB)
through which the elementary particles acquire their masses, and which 
necessitated the introduction of a new particle, known as the Higgs scalar~\cite{Higgs}.
But this mechanism has only recently come under experimental scrutiny,
with the discovery by the LHC collaborations ATLAS and CMS of a particle with properties
similar to those expected for the SM Higgs~\cite{cms_atlas}.
The particle is produced by colliding two protons ($pp$) and
detected, at present, mostly through its decay into two
photons ($\gamma \gamma$) and its decay into two $Z$ bosons ($ZZ$).
It turns out that the specific production mechanism can
sometimes be tagged and that it affects the detectability.
In the case of $\gamma \gamma$, results are known for all production mechanisms
combined, and also for production of the Higgs through
the fusion of two $Z$ or of two $W$ bosons, known collectively as vector boson fusion (vbf).

The field content of the SM has been determined by experiment.
Indeed, CP violation aside, the theory would have been viable if there were only
one charged lepton, one neutrino,
one up-quark, and one down-quark.
But three such families have been found, with increasing masses.
In the same fashion, there may be more than one scalar doublet,
and only experiment will tell. The two Higgs doublet model
was proposed by Lee~\cite{Lee:1973iz,review} in 1973 as a means of trying to explain
the matter-antimatter asymmetry in the universe. It is a very compelling
generalization of the SM, with a richer scalar spectrum: there would be three neutral 
scalars ($h$ and $H$, respectively the lightest and heaviest
CP-even scalars; and $A$, the pseudoscalar);
and a pair of conjugate charged scalars ($H^\pm$).
But having more than one scalar doublet also enriches the
vacuum structure. In the SM with one scalar doublet,
the potential admits only one minimum at tree-level,
up to gauge transformations. Nevertheless, quantum corrections
may induce vacuum metastability - in fact, the running of the 
scalar quartic coupling can be driven to negative values at 
very high energy scales, due to the largeness of the top 
quark Yukawa. This would correspond to the existence of a deeper minimum
of the potential, and ``ours" would be a false vacuum.  
Recent calculations have shown that, with a 
Higgs mass around 125 GeV, the scalar self-coupling would become negative well below
the Planck mass - depending on the uncertainties on the top quark pole mass
and on the strong coupling constant, this would occur between $10^{10}$ and
$10^{14}$ GeV~\cite{Degrassi:2012ry}.

In the 2HDM there is the possibility of metastable vacua even at tree-level. In fact,
the vacuum structure of the 2HDM is much richer than the SM's: for instance,
charge breaking (CB) and CP breaking minima can occur in the 2HDM, but not in the SM. 
It has been proven~\cite{Ferreira:2004yd,Barroso:2005sm} that no minima of
different natura can coexist in the 2HDM: if a charge and CP preserving
minimum exists, then a CB or CP stationary point is necessarily a saddle point, and
lies above it. And if a CP (or CB) minimum occurs, all other types of stationary points have
to be saddle points. Nonetheless, a different possibility occurs: the 2HDM can have
{\em two} minima which preserve CP, but break the standard electroweak gauge 
symmetries. And those two minima can be non-degenerate, corresponding to a completely
different mass spectrum for all elementary particles in each of them. And again,
we stress that this situation arises at tree-level. Radiative corrections will 
undoubtedly have relevant contributions to make, but the possibility of metastability 
in the 2HDM is potentially more serious than in the SM.

In fact, vacuum metastability raises a troubling possibility. If we were at present in 
the metastable vacuum, then the scalar field could in time decay into the real vacuum. 
Since in the deeper vacuum the scalar fields' vacuum expectation values have in general
very different values, tunneling to the true vacuum of the model would alter all particle
masses, with dramatic consequences for the entire universe. We call this situation a 
``panic vacuum". In this paper, we will consider one of the simplest versions of the
2HDM where this situation might occur - the Peccei-Quinn model. We will present the conditions 
that the parameters of the theory need to obey in order to avoid all tree-level metastability. And we will 
show that, remarkably, the current experimental results from the LHC  can already be used
to probe this question. In fact, they can already exclude, at the $2\sigma$ level, any possibility 
of panic vacua. We will also present an estimate of the lifetime of the false vacua, and show that 
most of them would have a tunneling time inferior to the age of the universe. Nonetheless, 
the LHC is already telling us that the Peccei-Quinn model, if it describes 
particle physics, has a vacuum which is completely stable at tree-level. 

\section{False vacua in the Peccei-Quinn model}

In this paper we will study one of the simplest possible 
versions of the 2HDM, the Peccei-Quinn model~\cite{PQ},~\cite{review}.
This model has a global $U(1)$ symmetry, such that the scalar fields transform as
$(\Phi_1, \Phi_2) \to (\Phi_1, e^{i \xi} \Phi_2)$. Extending this symmetry to the Yukawa sector, 
it is possible to make sure that the model possesses no tree-level favour changing neutral
currents (which would be present otherwise). It also considerably simplifies the
scalar potential, which is given by
\begin{eqnarray}
V_H
&=&
m_{11}^2 \Phi_1^\dagger \Phi_1 + m_{22}^2 \Phi_2^\dagger \Phi_2
- m_{12}^2 \left[  \Phi_1^\dagger \Phi_2 + \Phi_2^\dagger \Phi_1 \right]
\nonumber\\[6pt]
&+&
\tfrac{1}{2} \lambda_1 (\Phi_1^\dagger\Phi_1)^2
+ \tfrac{1}{2} \lambda_2 (\Phi_2^\dagger\Phi_2)^2
\nonumber\\[6pt]
&+&
\lambda_3 (\Phi_1^\dagger\Phi_1) (\Phi_2^\dagger\Phi_2)
+ \lambda_4 (\Phi_1^\dagger\Phi_2) (\Phi_2^\dagger\Phi_1),
\label{VH1}
\end{eqnarray}
where all parameters are real because CP invariance is imposed. Notice that we
have included a soft breaking term - the $m_{12}^2$ term - to prevent the
appearance of a massless pseudoscalar (an axion) when both doublets acquire
a vacuum expectation value (vev). 

In this model, spontaneous CP breaking cannot occur. And since the electromagnetic
symmetry is unbroken, this means that, after SSB, only the neutral components of the 
scalars will develop vevs. That is, 
$\langle \Phi_1^0 \rangle = v_1/\sqrt{2}$
and $\langle \Phi_2^0 \rangle = v_2/\sqrt{2}$,
which can be traded for $v = \sqrt{v_1^2 + v_2^2}$
and $\tan{\beta} = v_2/v_1$. Without loss of generality,
we can take $v_1$ and $v_2$ positive and $0 \leq \beta \leq \pi/2$.
The vacuum that our Universe is currently in has
$v = 2 m_W/g = 246~{\rm GeV} $ (related to the $W$ mass and the weak coupling $g$).
Now, it has been shown~\cite{ivanov,Barroso:2007rr} that the 2HDM can in fact have
at most {\em two} minima of this kind - the first with the vevs $v_1$ and $v_2$ 
defined above,  and the  second with vevs $v^\prime_1$ and $v^\prime_2$,
such that $\sqrt{{v^\prime_1}^2 + {v^\prime_2}^2} \neq v$. We are particularly interested
in the possibility - which we have called the  ``\textit{panic vacuum}'' - that the
vacuum we are living in is not the deepest one. The second set of vevs would thus
correspond to a lower potential, 
\begin{equation}
V_H(v^\prime_1,v^\prime_2) < V_H(v_1,v_2).
\label{panic}
\end{equation}
The vevs can be obtained by minimizing $V_H$, leading to two coupled cubic equations.
Given $m_{11}^2 \dots \lambda_4$, these can be solved numerically. Fortunately,
many interesting features of the vacuum can be determined without solving those equations explicitly.
In fact, in Ref.~\cite{ivanov} a general study of the conditions under which these two minima
arise was undertaken, and conditions for their existence, and relative depth, were established. 
These are re-derived and expanded in a parallel paper~\cite{us}. Applying the results of~\cite{ivanov}
to the specific case of the softly broken Peccei-Quinn model, 
the two minima can occur only if
\begin{eqnarray}
m_{11}^2 + k^2\, m_{22}^2
&<& 0,
\label{M_0}
\\
\sqrt[3]{x^2} + \sqrt[3]{y^2}
&\leq&  1,
\label{astroid}
\end{eqnarray}
where
\begin{eqnarray}
x
&=&
\frac{4\ k\ m_{12}^2}{
m_{11}^2 + k^2\, m_{22}^2}\,
\frac{\sqrt{\lambda_1 \lambda_2}}{
\lambda_{34} - \sqrt{\lambda_1 \lambda_2}},
\nonumber\\
y
&=&
\frac{m_{11}^2 - k^2\, m_{22}^2}{
m_{11}^2 + k^2\, m_{22}^2}\,
\frac{\sqrt{\lambda_1 \lambda_2} + \lambda_{34}}{
\sqrt{\lambda_1 \lambda_2} - \lambda_{34}},
\label{xANDy}
\end{eqnarray}
with
\begin{equation}
\lambda_{34} = \lambda_3 + \lambda_4
\hspace{5mm}
\textrm{and}
\hspace{5mm}
k = \sqrt[4]{\frac{\lambda_1}{\lambda_2}}.
\end{equation}
The limiting curve
$\sqrt[3]{x^2} + \sqrt[3]{y^2} =  1$
is known as an astroid. If either of conditions~\eqref{M_0}
or~\eqref{astroid} are not verified, then the potential has a single
minimum (which is of course the global minimum of the potential). 

However, in the region where two minima can coexist, one must investigate 
whether the minimum with vevs $v_1$, $v_2$ is, or isn't, the global minimum
of the model. It turns out~\cite{ivanov,us} that answering that question
ends up being very simple. All one has to do is to define the following  
quantity, which we call a discriminant,
\begin{equation}
D =
\left( m_{11}^2 - k^2 m_{22}^2 \right)
(\tan{\beta} - k).
\label{D}
\end{equation}
The discriminant is, of course, computed in our vacuum. 
Then, the following theorem holds:

\ 

{\em Our vacuum is the global minimum of the theory if and only if $D>0$.}

\

If, on the other hand, we find that $D<0$, then Eq.~\eqref{panic} holds and 
we live in a metastable state, and  our current $(v_1, v_2)$
solution is the ``\textit{panic vacuum}''.

Conversely, if there is only one minimum, or even if there are two minima
-- Eqs.~\eqref{M_0} and \eqref{astroid} hold -- but $D>0$ in the region where it is defined,
then $(v_1,v_2)$ corresponds to the global minimum of the theory,
and our vacuum is stable. This means that, to ascertain the nature of the Peccei-Quinn model
vacuum, all one truly needs is to verify the value of the discriminant $D$. 

Problems related to ours are sometimes tackled by following the evolution of the various minima as
the temperature ($T$) of the Universe is decreased, from its inception to the present day.
This involves estimates of the finite temperature contributions to the effective potential,
and hinges on a variety of assumptions. Some recent work within the 2HDM can be found in
Ref.~\cite{finite-T}. It is also true that, even at $T=0$, the nature of the vacuum changes
as we increase the energy scale above 10~TeV. As we mentioned earlier, this occurs even in the SM,
due to loop corrections to $V_H$, and might ultimately point to a more complete theory
at high energies, encompassing the SM~\cite{Degrassi:2012ry}. This is not our concern here, since we 
are worried about a problem that affects us right now, at $T=0$ and small energies. One might expect 
that radiative corrections will be important in the 2HDM, as they are in the SM. In fact, they might 
even raise the possibility of further deeper minima existing. But the issue in the 2HDM is more
immediate, since both minima exist at tree level and $T = 0$.

\section{Panic vacua and LHC results}

We will now show that measurements in particle physics experiments
can be used to reconstruct the shape of the present day
($T=0$) tree-level scalar potential, and,
thus, ascertain as to the putative presence of a lower lying minimum.
Even more striking, although there are not,
as of yet, enough measurements to determine all the seven
$m_{11}^2 \dots \lambda_4$ parameters of $V_H$ - or even any indication
of the existence of more than one scalar -  the experiments at the LHC are already 
close to ruling out the \textit{panic vacuum} situation of
Eq.~\eqref{panic}. In fact, they strongly disfavour the possibility of
a panic vacuum at the $2\sigma$ level. 

If the soflty broken Peccei-Quinn model describes Nature,
and the three neutral scalars and conjugate pair of charged
scalars are detected, then parameters of the scalar potential can be 
reproduced from the following relations,
\begin{eqnarray}
m_{12}^2
&=&
m_A^2\, s_\beta c_\beta,
\nonumber\\
\lambda_1
&=&
\frac{- s_\beta^2 m_A^2 + c_\alpha^2 m_H^2 + s_\alpha^2 m_h^2}{v^2 c_\beta^2},
\nonumber\\
\lambda_2
&=&
\frac{- c_\beta^2 m_A^2 + s_\alpha^2 m_H^2 + c_\alpha^2 m_h^2}{v^2 s_\beta^2},
\nonumber\\
\lambda_3
&=&
\frac{2 m_{H^\pm}^2 - m_A^2}{v^2} +
\frac{s_{2 \alpha} (m_H^2 - m_h^2)}{v^2 s_{2 \beta}},
\nonumber\\
\lambda_4
&=&
\frac{2 (m_A^2 - m_{H^\pm}^2)}{v^2},
\label{couplings}
\end{eqnarray}
where $c_\theta = \cos{\theta}$ and $s_\theta = \sin{\theta}$,
for any needed angle $\theta$,
and the $m$'s are the masses of the various scalars.
The $m_{11}^2$ and $m_{22}^2$ parameters are obtained by solving
the stationarity conditions
$\partial V_H/(\partial v_1) = 0$ and $\partial V_H/(\partial v_2) = 0$.
The angle $\alpha$ parametrizes the mixing
\begin{eqnarray}
h &=&
\textrm{Re}(\Phi_1^0)\, s_\alpha - \textrm{Re}(\Phi_2^0)\, c_\alpha,
\nonumber\\
H &=&
- \textrm{Re}(\Phi_1^0)\, c_\alpha - \textrm{Re}(\Phi_2^0)\, s_\alpha,
\end{eqnarray}
and can be taken between $-\pi/2$ and $\pi/2$.
Eventually, the angles $\alpha$ and $\beta$ will be determined
by probing the couplings of the scalar particles to the
fermions. Usually, two types of couplings to fermions are considered
in theories with the $U(1)$ symmetry. We thus consider two different
models, with the same scalar potential but different scalar-fermion interactions.
In model Type-I, the fermions transform under the Peccei-Quinn
symmetry in such a manner that all fermions couple only to the $\Phi_2$ doublet.
But in model Type-II, the fermion's transformation laws are chosen such that
up-quarks couple to $\Phi_2$, while down-quarks and charged leptons couple to $\Phi_1$.
Both of these types of models have interesting phenomenologies, and are constrained
in different manners by existing data, mostly coming from B-physics experiments~\cite{Mahm}.

It could happen that the panic vacua solutions could only occur for uninteresting 
regions of parameter space - for instance, if they could only occur for, say a Higgs 
mass of 50 GeV, we could dismiss metastable vacua in the Peccei-Quinn model as being
only a theoretical curiosity. In order to probe the interest of the panic vacua solutions 
we have therefore performed a vast scan of the parameter space of the model and 
verified where those solutions arose. 
We have  randomly generated $180000$ points in the parameter space - since $v$ and $m_h$ are fixed, 
each ``point"  corresponds to a combination of five different parameters. Ours is therefore
a 5-dimensional parameter space, and all plots we present are to be understood as ``slices" of that 
higher-dimensional space. In our scan, we set $v = 246\, \textrm{GeV}$ and $m_h = 125\, \textrm{GeV}$,
and varied the remaining parameters in the ranges 
$m_H \in [125, 900]\, \textrm{GeV}$,
$m_A \in [10, 900]\, \textrm{GeV}$,
$m_{H^\pm} \in [90, 900]\, \textrm{GeV}$,
and $ - \pi/2 \leq \alpha \leq \pi/2$.
In addition, due to experimental bounds coming from $B$-physics~\cite{Mahm}, 
we have kept $\tan{\beta} > 1$.
The corresponding couplings in Eqs.~\eqref{couplings}
have been forced to comply with a scalar potential bounded from below~\cite{Deshpande:1977rw},
to satisfy tree-level unitarity~\cite{Kanemura:1993hm,Akeroyd:2000wc}, and to be consistent with constraints
from the electroweak precision observables $S$, $T$ and $U$~\cite{gfitter1,gfitter2}.

\begin{figure}[htb]
\epsfysize=6.5cm
\centerline{\epsfbox{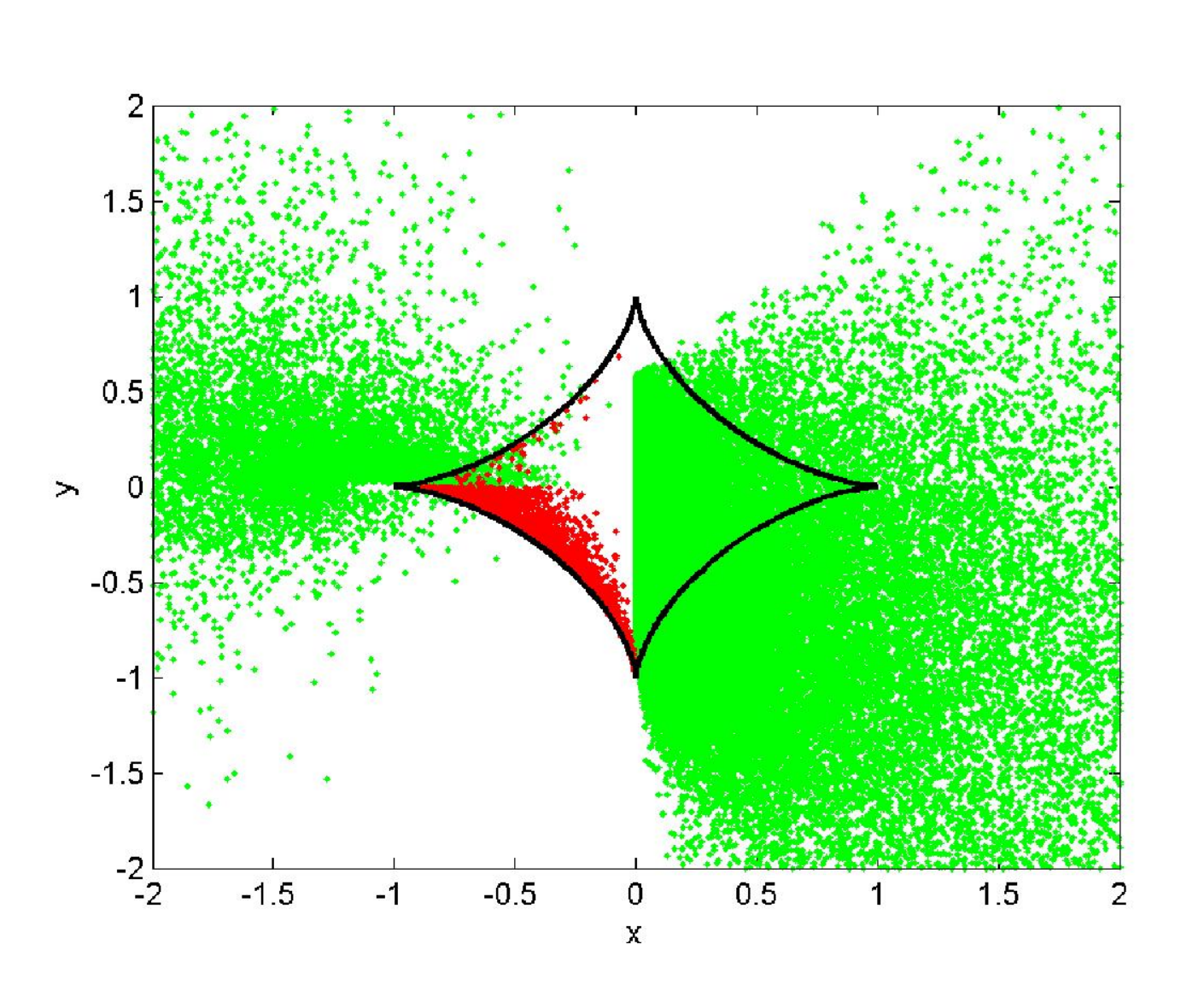}}
\caption{Placement in the $(x, y)$ plane of the points generated.
Points inside the astroid (solid lines) can have two minima;
those in red/dark-gray correspond to the panic vacuum.}
\label{fig1}
\end{figure}

How often can both minima exist in this model? To verify this, we have computed the $x$ and $y$ 
variables of Eqs.~\eqref{xANDy} for each combination of parameters in the potential and plot the value
of Eq.~\eqref{astroid} in Fig.~\ref{fig1}. There we 
show the generated points in the
$(x, y)$ plane and in solid black lines the astroid which delimits the
region where dual minima may exist.
Points inside (outside) the astroid  satisfy (do not satisfy)
Eq.~\eqref{astroid} and correspond to potentials which do
(do not) have two minima.
Green/light-gray points inside the astroid
have $D > 0$ and exhibit two minima,
with our current vacuum being the global minimum.
This is a safe situation, no tunneling from our vacuum to another can occur
(at tree level).
Red/dark-gray points inside the astroid have $D < 0$, exhibit two minima,
but correspond to the panic situation where our current vacuum lies above
the other one. We have checked by explicit computation
that the red/dark-gray points satisfy Eq.~\eqref{panic},
while the green/light-gray points inside the astroid do not. As we see from 
Fig.~\ref{fig1}, the occurrence of two minima is not a rare event in the 2HDM - a ``blind"
scan of parameters encounters many such minima, and among them the panic vacua aren't rare
either.

In order to study the panic vacuum in detail,
we have generated two new sets of points
where Eqs.~\eqref{M_0}-\eqref{astroid} hold,
and $D < 0$. The data sample for model Type-I has over $100000$
points. In the Type-II model we have a further constraint -  the
charged scalar mass has to be such that
$m_{H^\pm} > 340\, \textrm{GeV}$. This arises from $b \to s \gamma$ 
measurements, and is almost independent of $\tan{\beta}$~\cite{Mahm}.
For model Type-II 
we have thus generated $58000$ points
obeying this further constraint, as well as all the other mentioned previously.

The generated points are used to calculate $R_f$,
defined as the number of events predicted in the model
for the process $pp \rightarrow h \rightarrow f$,
divided by the prediction
obtained in the SM for the same final state $f$.
In other words, 
\begin{equation}
R_f \,=\, \frac{\sigma^{2HDM} (pp\rightarrow h) \,BR^{2HDM}(h \rightarrow f)}
{\sigma^{SM} (pp\rightarrow h) \,BR^{SM}(h \rightarrow f)},
\end{equation}
with production cross sections $\sigma$ and the branching ratios (BR) of the Higgs 
boson considered for both models. We have considered all Higgs production mechanisms
which are at work at LHC: gluon-gluon (gg) fusion; vector boson fusion (vbf);
associated production of a Higgs and a vector boson or a $t\bar{t}$ pair; and $b\bar{b}$
fusion. The two LHC observables which give more precise results for Higgs physics at the 
moment are the $R$ ratios to two photons or two $Z$ bosons. 
 Their experimental bounds at the $1 \sigma$
level are~\cite{average}
$R_{ZZ} = 0.93 \pm 0.28$,
$R_{\gamma\gamma} = 1.66 \pm 0.33$,
summing over all production mechanisms.

\begin{figure}[htb]
\epsfysize=6.5cm
\centerline{\epsfbox{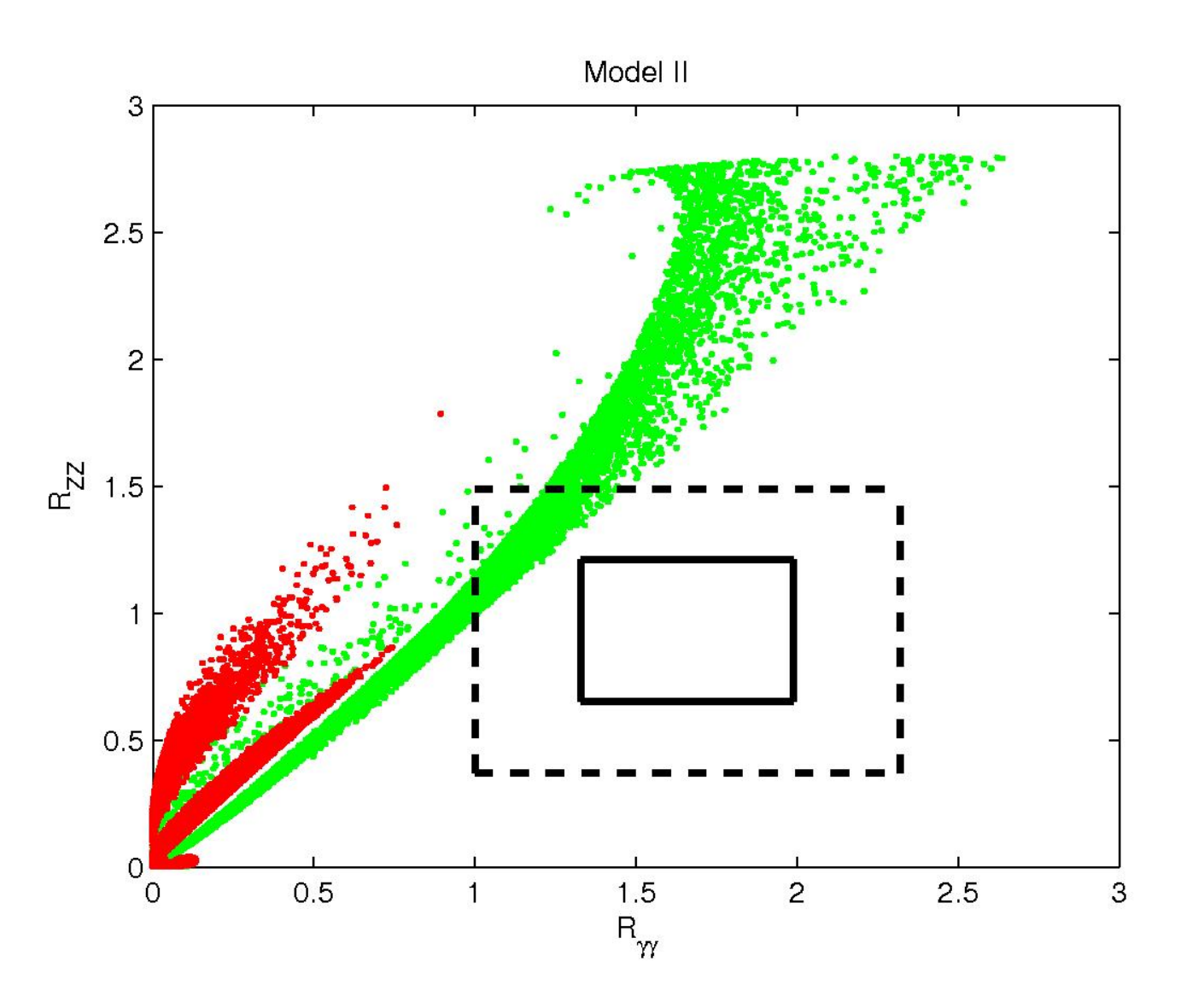}}
\caption{Predictions from Type-II
in the $(R_{\gamma\gamma}, R_{ZZ})$ plane for
panic vacuum points (in red/dark-gray)
and for non-panic points (in green/light-gray).
Also shown are the $1 \sigma$ (solid line) and
$2 \sigma$ (dashed line) experimental bounds.}
\label{fig2}
\end{figure}

\begin{figure}[htb]
\epsfysize=6.5cm
\centerline{\epsfbox{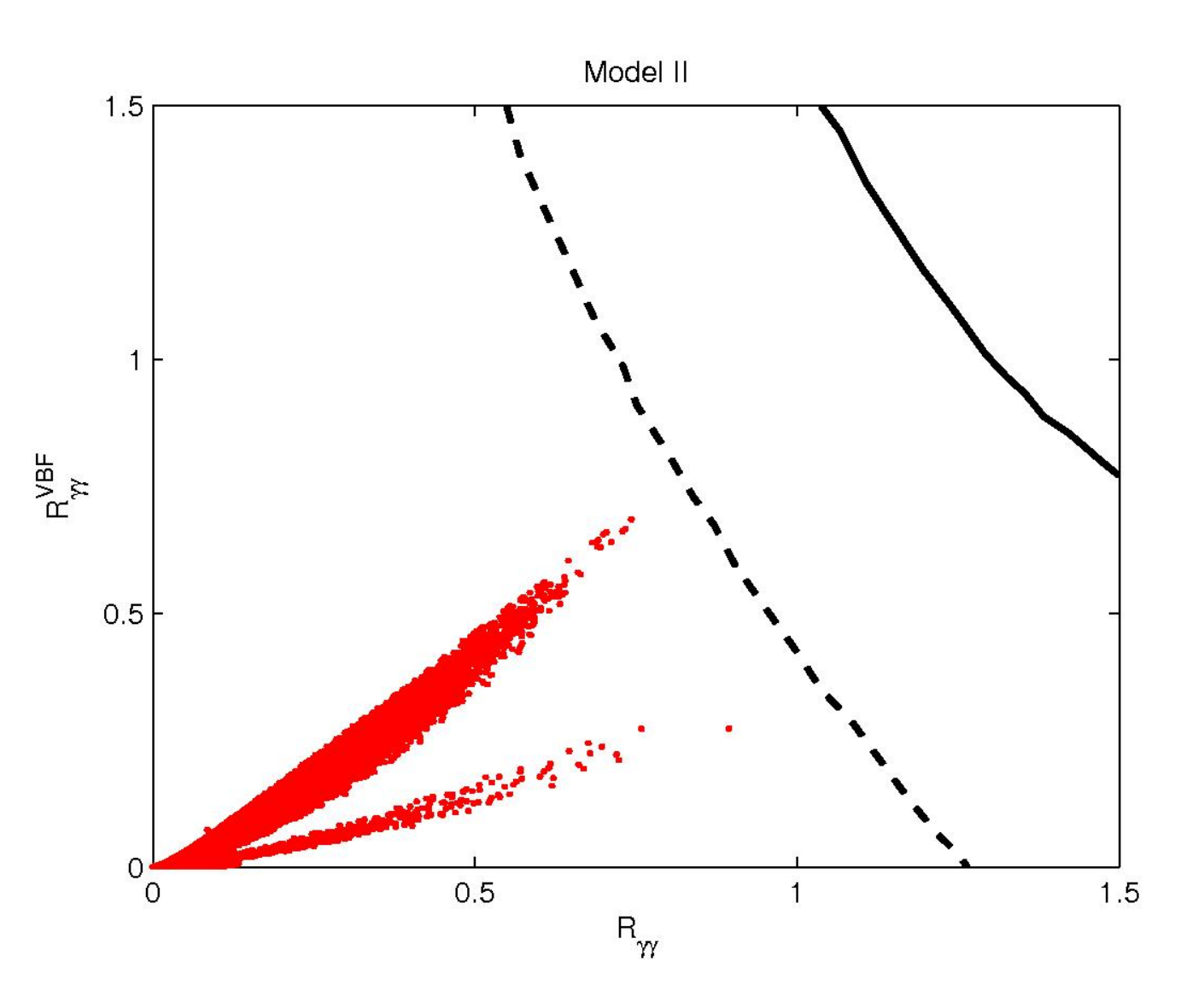}}
\caption{Prediction for panic vacuum points in Type-II
in the $(R_{\gamma\gamma}, R^{\rm vbf}_{\gamma\gamma})$ plane.
They lie outside the $1 \sigma$ (solid line) and $2 \sigma$
(dashed line) ellipse bounds in Fig.~12 of Ref.~\cite{elipse}.}
\label{fig3}
\end{figure}

The results we obtain for Type-II are shown in
Figs.~\ref{fig2} and \ref{fig3}.
Fig.~\ref{fig3} also includes the $1 \sigma$ (solid line)
and $2 \sigma$ (dashed line) bounds coming from the
ellipse in Fig.~12 of Ref.~\cite{elipse}.
While model Type-II  points (green/light-gray) for which our current
vacuum coincides with the global minimum are consistent
with experiment,
we find that model Type-II panic points (red/dark-gray)
are excluded at least at the $2\sigma$ level by both $ZZ$
and by $(R_{\gamma\gamma}, R^{\rm vbf}_{\gamma\gamma})$.
Since our parameter space is 5-dimensional, it is very hard
to find a simple-to-grasp explanation as to why the panic points,
which obey $D<0$ as well as Eqs.~\eqref{M_0} and~\eqref{astroid},
~should concentrate as they do, for lower values of $R_{\gamma\gamma}$.
But the numerical scan shows they do, and as such are strongly disfavoured
by LHC data.

\begin{figure}[htb]
\epsfysize=6.5cm
\centerline{\epsfbox{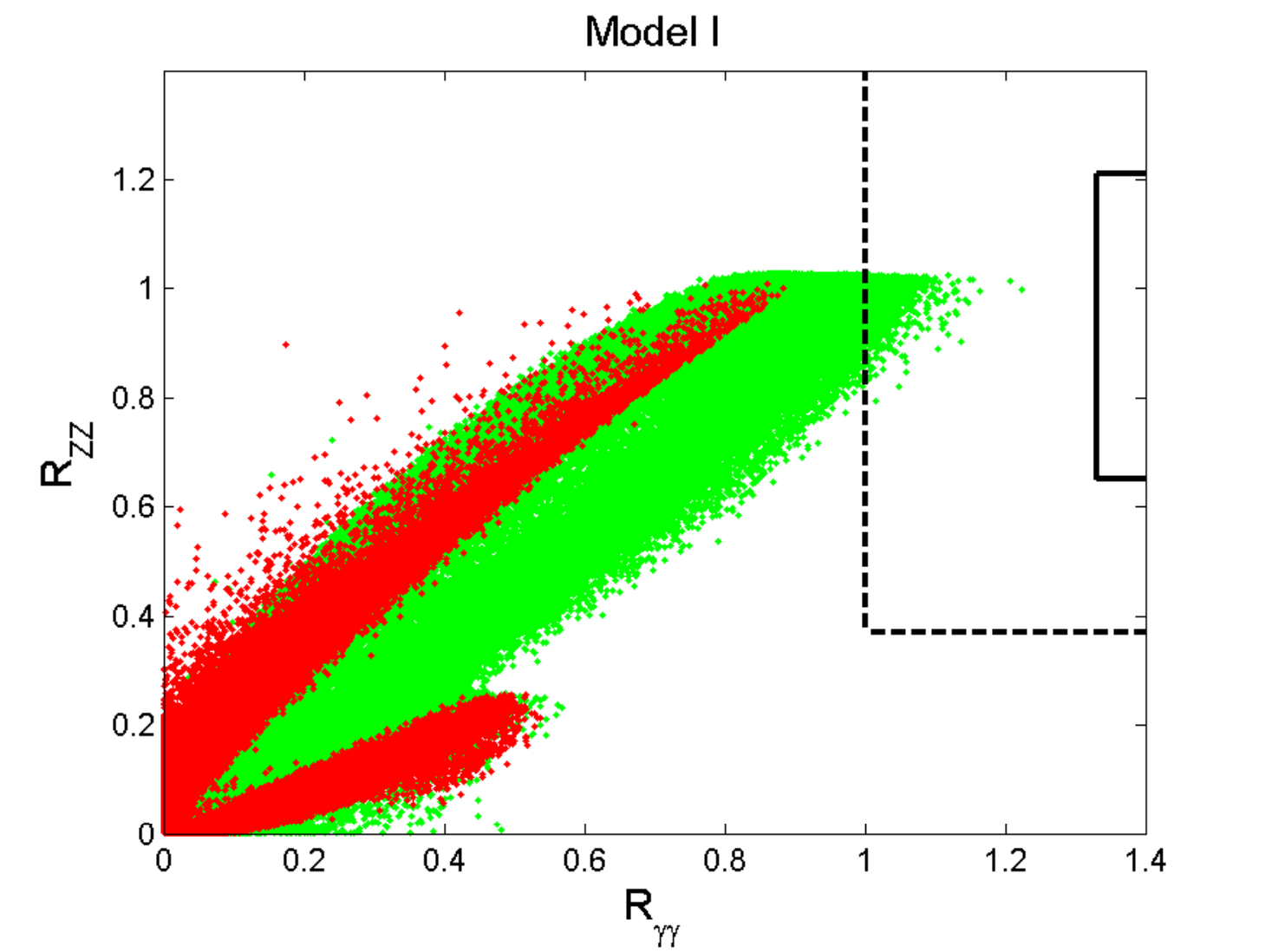}}
\caption{Predictions from Type-I
in the $(R_{\gamma\gamma}, R_{ZZ})$ plane for
panic vacuum points (in red/dark-gray)
and for non-panic points (in green/light-gray).
Also shown are the $1 \sigma$ (solid line) and
$2 \sigma$ (dashed line) experimental bounds.}
\label{fig4}
\end{figure}

The results we obtain for model Type-I are shown in
the $(R_{\gamma\gamma}, R_{ZZ})$ plane of
Fig.~\ref{fig4}.
We find that $R_{\gamma\gamma} \lesssim 0.88$,
$R_{ZZ} \lesssim 1$,
and the panic points lie outside the
$2 \sigma$ $R_{\gamma\gamma}$ experimental band.
In contrast,
the Type-I panic points we have generated in the
$(R_{\gamma\gamma}, R^{\rm vbf}_{\gamma\gamma})$ plane
lie outside the $1 \sigma$ (elliptical) band,
but inside the $2 \sigma$ band.
We could have $R_{\gamma\gamma}^{\rm vbf} > 1$,
but only for values of $R_{\gamma\gamma} < 0.6$.
Thus, we cannot reach the SM values,
$R_{\gamma\gamma} = R_{ZZ} = R_{\gamma\gamma}^{\rm vbf} = 1$,
and the measurements of $R_{\gamma\gamma}$
are barely consistent with our results for Type-I.

\section{False vacuum lifetime estimates}

Thus far, we have taken the view that, regardless of
how one estimates the Universe's evolution, the possibility
that we are now (or not) in a metastable vacuum
can be ascertained by measurements in particle physics experiments. 
But it is interesting to have an estimate of what cosmology has to say about
the likelihood of a panic vacuum in our model. In fact, one usually takes the 
point of view that, if the tunneling time between the false vacuum and
the true one is larger than the current age of universe, then the parameters
of the potential which produce said false vacuum are perfectly acceptable. 
In fact, such a situation would be indistinguishable from our vacuum 
being the model's global minimum. In a theory with a single scalar, the
lifetime of the false vacuum is estimated calculating the bounce trajectory 
between both minima~\cite{cole}. That trajectory passes through the  maximum between both minima. 
The estimate of the vacuum lifetime is however very difficult in the 2HDM, given that
there are many field directions present. In fact,
even removing the would-be Goldstone bosons, the scalar potential is still
a five-variable function. Thus, the path from one minimum to the other can be much
``shorter" than what one would expect. In fact, it doesn't even need to pass by the potential's
maximum, it can pass by a saddle point, or in fact avoid any stationary point
whatsoever~\cite{ruba}. We will follow the standard procedure and take
the thin wall approximation of Refs.~\cite{cole,ruba,sov}. We therefore estimate the
decay width of the false vacuum per unit volume as the exponential of
\begin{equation}
-B\,=\,-\,\frac{2^{11} \pi^2}{3\lambda}\,\left(\frac{\delta}{\epsilon}\right)^3\,.
\end{equation}
In this formula
$\delta$ is the height of the barrier at the saddle point
between the two minima, relative to the highest minimum - in this way we are 
trying be be conservative and taking the most unfavourable estimate, since this saddle point is 
actually {\em lower} than the potential's maximum, thus quickening the tunneling time. 
As for $\epsilon$, it is the difference of the values
of the potential calculated in the two minima. To further make sure our estimate is the most conservative one, 
we take $\lambda$ as the largest of the quartic couplings in the scalar potential,
$\lambda = max(|\lambda_k|)$ ($k = 1 \ldots 5$). Having $\delta/\epsilon$ larger than order
one is already sufficient to stabilize the Universe; for
example, Ref.~\cite{carena} quotes $B > 400$. Thus, large values of
$\delta/\epsilon$ correspond to a tunneling time to the deeper vacuum
larger than the age of the Universe.

In order to perform this lifetime estimate, we generated a separate sample of points, 
consisting only of panic vacua points, and for which we had all information concerning both vacua: 
namely, both sets of vevs $\{v_1\,,\,v_2\}$ and $\{v^\prime_1\,,\,v^\prime_2\}$. This is
simple to achieve in a numerical way, although it would be impossible to present analytical
expressions for both sets of vevs. It is further possible, employing the Lagrange multiplier
formalism used in~\cite{ivanov}, to discover the remaining stationary points of the
potential, and thus find the closest saddle point to the panic vacuum. In this manner
we can obtain the values of both $\delta$ and $\epsilon$. 

With that calculation, we have been able to show that, out of all the points 
shown in Fig.~\ref{fig1}, only about 5\% have estimated lifetimes which are
larger than the current age of the universe. They are located close to the axis
of the astroid. In contrast, points with small values of $\delta/\epsilon$ correspond to
a tunneling time to the deeper vacuum smaller than the
age of the Universe. These points must be excluded 
from the parameter space on cosmological grounds, because
they would correspond to a situation in which the
Universe would already have decayed away into the global
minimum. The lifetime estimate we performed thus shows that the vast majority of the
panic vacua one finds are indeed unacceptable choices of parameters, since they predict
a vacuum which would have decayed long ago. However, we must stress our point: 
although these cosmological estimates preclude a good portion (but not all) of the
panic vaccua, they hinge on the thin wall approximation that breaks down precisely when 
$\delta/\epsilon$ is very small. One could argue that  a more accurate (and much more complicated)
calculation of the lifetime could significantly shift its numerical values. As such, we cannot,
based on this simple calculation, be certain which part of the astroid can be disregarded
on cosmological grounds, and which is cosmologically viable. 
In contrast, current LHC bounds allow us to already exclude
all panic vacuum points, {\em regardless of any lifetime estimate}. The beautiful thing in this
model is that the lifetime estimate ends up being unnecessary, since LHC data already 
tells us that the Peccei-Quinn vacuum is a global one.

\section{Conclusions}

We have studied the vacuum structure of a
2HDM with a softly broken $U(1)$ symmetry,
looking out for situations
where our current vacuum has a larger energy
density than the true global
minimum of the scalar potential.
We stress that the exclusion of panic vacuua should be addressed
in models with extended Higgs sectors,
where it should be implemented as an extra constraint on the
theory's parameter space. In fact, the conditions for panic vacua we present in
this work are a special case of those studied in ref.~\cite{us}. The discriminant
we presented in this paper, Eq.~\eqref{D}, is the only quantity we need to compute to
ascertain whether a vacuum of the softly broken Peccei-Quin model is, or isn't the
global minimum of the potential. 

What we have found is that, in this model, the current
LHC results allow us to conclude, at the $2\sigma$ level, that it is impossible that we
are currently living in the higher minimum - even without evidence of any extra scalars, the 
LHC already provides us information about the nature of the 2HDM vacuum, which is quite remarkable. 
For many of the parameter points we found in which panic vacua occurred - that is, where our
vacuum would be the highest minimum of the potential - a standard estimation of the tunneling
time to the lower minimum tells us that we could no longer be there. That is, most of the
panic vacua we found, characterized by $D < 0$, are (according to standard vacuum lifetime 
estimates) phenomenologically unacceptable, as they predict a vacuum which would have decayed long
ago. Hence, such parameters could, on cosmological grounds, be excluded from the model's
parameter space. But an interesting aspect of our analysis is that we need not even worry
about vacuum lifetimes - the LHC data already enables us to exclude panic vacuum solutions. 

Our conclusion is that, within the softly broken Peccei-Quinn model,
the recent LHC experiments are crucial in excluding
the panic vacuum in model Type-II and strongly disfavoring it
in model Type-I.

\begin{acknowledgments}
The works of A.B., P.M.F. and R.S. are supported in part by the Portuguese
\textit{Funda\c{c}\~{a}o para a Ci\^{e}ncia e a Tecnologia} (FCT)
under contract PTDC/FIS/117951/2010, by FP7 Reintegration Grant, number PERG08-GA-2010-277025,
and by PEst-OE/FIS/UI0618/2011.
The work of J.P.S. is funded by FCT through the projects
CERN/FP/109305/2009 and  U777-Plurianual,
and by the EU RTN project Marie Curie: PITN-GA-2009-237920.
\end{acknowledgments}

\end{document}